# An Optimized Analogy-Based Project Effort Estimation

Mohammad Azzeh
Faculty of Information Technology
Applied Science University
Amman, Jordan POBOX 166
`m.y.azzeh@asu.edu.jo`

Yousef Elsheikh
Faculty of Information Technology
Applied Science University
Amman, Jordan
`y_elsheikh@asu.edu.jo,`

Marwan Alseid
Faculty of Information Technology
Applied Science University
Amman, Jordan
`m_alseid@asu.edu.jo`

**Abstract.**

Despite the predictive performance of Analogy-Based Estimation (ABE) in generating better effort estimates, there is no consensus on how to predict the best number of analogies, and which adjustment technique produces better estimates. This paper proposes a new adjusted ABE model based on optimizing and approximating complex relationships between features and reflects that approximation on the final estimate. The results show that the predictive performance of ABE has noticeably been improved, and the number of analogies was remarkably variable for each test project.

**Keywords**: Cost Estimation; Effort Estimation by Analogy; Bees Optimization Algorithm

## 1. Introduction

Analogy-Based Estimation (ABE) has preserved popularity within software engineering research community because of its outstanding performance in prediction when different data types are used [1, 15]. The idea behind this method is rather simple such that the new project's effort can be estimated by reusing efforts about similar, already documented projects in a dataset, where in a first step one has to identify similar projects which contain the useful predictions [15]. The predictive performance of ABE relies significantly on the choice of two interrelated parameters: number of nearest analogies and adjustment strategy [8]. The goal of using adjustment in ABE is twofold: (1) minimizing the difference between a new project and its nearest analogies, and (2) producing more successful estimates in comparison to original ABE [2]. If the researchers read the literature on ABE, they will encounter large number of ABE models that use variety of adjustment strategies. Those strategies suffer from common problems such as they are not able to produces stable results when applied in different contexts as well as they use fixed number of analogies for the whole dataset [1]. Using fixed number of analogies has been proven to be unsuccessful in many situations because it depends heavily on expert opinion and requires extensive experimentation to identify the best *k* value, which might not be predictive for individual projects [2].

The aim of this work is therefore to propose a new method based on Artificial Bees Algorithm (BA) [14] to adjust ABE by optimizing the feature similarity coefficients that minimizes difference between new project and its nearest projects, and predicting the best *k* number of nearest analogies. The paper is structured as follows: Section 2 introduces an overview to ABE and adjustment methods. Section 3 presents the proposed adjustment method. Section 4 presents research methodology. Section 5 shows obtained results. Finally the paper ends with our conclusions.

## 2. Related Works

ABE method generates new prediction based on assumption that similar projects with respect to features description have similar efforts [8, 15]. Adjustment is a part of ABE that attempts to minimize the difference between new observation ($\hat{e}_i$) and each nearest similar observation ($e_i$), then reflects that difference on the derived solution in order to obtain better solution ($e_t$). Consequentially, all adjusted solutions are aggregated using simple statistical methods such as mean ($e_t = k^{-1}\sum_{i=1}^{k}\hat{e}_i$). In previous study [18] we investigated the performance of BA on adjusting ABE and finding best *k* value for the whole dataset. This model showed some improvements on the accuracy, but on the other side it did not solve the problem of predicting the best *k* value for each individual project. In addition the solution space of BA was a challenge because there was only one common weights for all nearest analogies. The used optimization criteria (i.e. *MMRE*) was problematic because it was proven to be biased towards underestimation. For all these reason and since we need to compare our proposed model with validated and replicated models, we excluded this model from comparison later in this paper. This paper thereby attempts to solve abovementioned limitations.

In literature there is a significant number of adjustment methods that have been documented and replicated in previous studies. Therefore we selected and summarized only the most widely used strategies. Walkerden and Jeffery proposed Linear Size Adjustment (LSE) [16] based on the size extrapolation. Mendes et al. [12] proposed Multiple Linear Feature Extrapolation (MLFE) to include all related size features. Jorgenson et al. [6] proposed Regression Towards the Mean (RTM) to adjust projects based on their productivity values. Chiu and Huang [4] proposed another adjustment based on Genetic Algorithm (GA) to optimize the coefficient αj for each feature distance based on minimizing performance measure. Recently, Li et al. [10] proposed the use of Neural Network (NN) to learn the difference between projects and reflects the difference on the final estimate. Further details about these methods and their functions can be found in [1].

Indeed, the most important questions to consider when to use such methods is how to predict the best number of nearest analogies (*k*). In recent years various approaches have been proposed to specify this number such as: 1) fixed number selection (i.e. *k*=1, 2, 3…etc) as in studies of [7, 11, 12, 16], 2) Dynamic selection based on clustering as in study of [2, 18]. 3) Similarity threshold based selection as in studies of [5, 9]. Generally, these studies except [2] use the same *k* value for all projects in the dataset which does not necessarily produce best performance for each individual project. On the other hand, the certain problem with [2] is that it does not include adjustment method but it predicts the best *k* value based on the structure of dataset.

3. **The Proposed Method** (OABE)

The proposed adjustment method starts with Bees Algorithm in order to find out, for each project: (1) the feature weights (*w*), and (2) the best *k* number of nearest analogies that minimize mean absolute error. The search space of BA can be seen as a set of *n* weight matrixes where the size of each matrix (i.e. solution) is $k \times m$. That means each possible solution contains weight matrix with dimension equivalent to the number of analogies (*k*) and number of features (*m*) as shown in Figure 1. The number of rows (i.e. *k*) and weight values are initially generated by random. Each row represents weights for one selected analogy and accordingly $\sum_{j=1}^{m} w_j = 1$. In each run the algorithm selects the top *k* nearest analogies based on the number of *k* weights in the search space. Then each selected analogy is adjusted with corresponding weights taken from the matrix *w* as shown Eq.1. The algorithm continues searching until the value of Mean Error (i.e. $MR = k^{-1} \sum_{j=1}^{k} \Delta_{ij}$) between new project and its *k* analogies is minimized. The optimized *k* value and weight matrix are then applied to Eqs. 1, 2 and 3 to generate new estimate. The new integration between ABE with BA will be called Optimized Analogy Based Estimation (hereafter OABE).

$$w = \begin{bmatrix} w_{11} & w_{12} & \cdots & w_{1m} \\ w_{21} & w_{22} & \cdots & w_{2m} \\ \cdots & \cdots & \cdots & \cdots \\ w_{k1} & w_{k2} & \cdots & w_{km} \end{bmatrix}$$

Fig. 1. Weight Matrix for one solution in the search space

$$\Delta_{ij} = \frac{1}{m} \sum_{j=1}^{m} w_{ij} \times (f_{tj} - f_{ij}) \qquad (1)$$

$$\hat{e}_i = e_i + \Delta_{ij} \qquad (2)$$

$$e_t = \frac{\sum_{i=1}^{k}(k+1-r_i)\times \hat{e}_i}{\sum_{i=1}^{k} j} \qquad (3)$$

The setting parameters for AB have been found after performing sensitivity analysis on the employed datasets to see the appropriate values. Table 1 shows BA parameters, their abbreviations and initial values used in this study. Below we briefly describe the process of BA in finding best *k* values and the corresponding weights for each new project. The algorithm starts with an initial set of weight matrixes generated after randomly initializing *k* for each matrix. The solutions are assessed and sorted in ascending order after they are being evaluated based on *MR*. The best from 1 to *b* solutions are being selected for neighborhood search for better solutions, and form new patch. Similarly, a number of bees (*nsp*) are also recruited for each solution ranked from *b+1* to *u*, to search in the neighborhood. The best solution in each patch will replace the old best solution in that patch and the remaining bees will be replaced randomly with other solutions. The algorithm continues searching in the neighborhood of the selected sites, recruiting more bees to search near to the best sites which may have promising solutions. These steps are repeated until the criterion of stop (minimum *MR*) is met or the number of iteration has finished.

Table 1. BA parameters

| Parameter | Description | Value |
|---|---|---|
| *q* | dimension of solution | (number of features +1) |
| *n* | represents size of initial solutions | 100 |
| *u* | number of sites selected out of *n* visited sites | 20 |
| *b* | number of best sites out of *s* selected sites | 10 |
| *nep* | number of bees recruited for best *b* sites | 30 |
| *nsp* | Number of bees recruited for the other selected sites | 20 |
| *ngh* | initial size of patches (*ngh*) | 0.05 |

## 4. Methodology
### 4.1. Datasets

The proposed OABE model has been validated over 8 software effort estimation datasets come from companies of different industrial sectors [3]. The datasets characteristics are provided in Table 2 which shows that the datasets are strongly positively skewed indicating many small projects and a limited number of outliers. It is important to note that all continuous features have been scaled and all observation with missing values are excluded.

**Table 2.** Descriptive statistics of the datasets

| Dataset | Feature | Size | Effort Data | | | | | |
|---|---|---|---|---|---|---|---|---|
| | | | unit | Min | max | mean | median | skew |
| Albrecht | 7 | 24 | months | 1 | 105 | 22 | 12 | 2.2 |
| Kemerer | 7 | 15 | months | 23.2 | 1107.3 | 219.2 | 130.3 | 2.76 |
| Nasa | 3 | 18 | months | 5 | 138.3 | 49.47 | 26.5 | 0.57 |
| Desharnais | 12 | 77 | hours | 546 | 23940 | 5046 | 3647 | 2.0 |
| COCOMO | 17 | 63 | months | 6 | 11400 | 683 | 98 | 4.4 |
| China | 18 | 499 | hours | 26 | 54620 | 3921 | 1829 | 3.92 |
| Maxwell | 27 | 62 | hours | 583 | 63694 | 8223.2 | 5189.5 | 3.26 |
| Telecom | 3 | 18 | months | 23.54 | 1115.5 | 284.33 | 222.53 | 1.78 |

## 4.2. Performance Measures

A key question to any estimation model is whether the predications are accurate, the difference between the actual effort ($e_i$) and the predicted effort ($\hat{e}_i$) should be as small as possible because large deviation will have opposite effect on the development progress of the new software project [13]. This section describes several performance measures used in this research as shown in Table 3. Although some measures such as *MMRE*, *MMER* have been criticized as biased to under and over estimations, we insist to use them because they are widely used in commenting on the success of predictions [13]. Interpreting these error measures without any statistical test can lead to conclusion instability, therefore we used *win-tie-loss* algorithm [8] to compare the performance of OABE to other estimation methods. We first check if two methods $method_i$; $method_j$ are statistically different according to the Wilcoxon test. If so, we update $win_i$; $win_j$ and $loss_i$; $loss_j$ after checking which one is better according to the performance measure at hand; otherwise we increase $tie_i$ and $tie_j$. The performance measures used here are *MRE, MMRE, MdMRE, MMER, MBER* and $Pred_{25}$. Algorithm 1 illustrates the win-tie-loss algorithm [8].

**Algorithm 1**. Pseudocode of win-tie-loss algorithm between $method_i$ and $method_j$ based on performance measure E [8]
```
 1: Win_i=0,tie_i=0,loss_i=0
 2: Win_j=0,tie_j=0;loss_j=0
 3: if Wilcoxon (MRE(method_i), MRE(method_j), 95) says they are the same then
 4:     tie_i = tie_i + 1;
 5:     tie_j = tie_j + 1;
 6: else
 7:     if better(E(method_i), E(method_j)) then
 8:         win_i = win_i + 1
 9:         loss_j = loss_j + 1
10:     else
11:         win_j = win_j + 1
12:         loss_i = loss_i + 1
13:     end if
14: end if
```

Also, the Bonferroni-Dunn test [17] is used to perform multiple comparisons for different models based on the absolute error to check whether there are differences in population rank means among more than populations.

Table 3. Summary of Performance Measures

| | |
|---|---|
| Magnitude Relative Error | $MRE = \dfrac{|e_i - \hat{e}_i|}{e_i}$ |
| Mean Magnitude Relative Error | $MMRE = N^{-1} \sum_i MRE_i$ |
| Medina Magnitude Relative Error | $MdMRE = median_i(MRE_i)$ |
| Mean Magnitude of Error Relative to the estimate | $MMER = N^{-1} \sum_i \dfrac{|e_i - \hat{e}_i|}{\hat{e}_i}$ |
| Mean Balanced Error (*MBRE*) | $MBER = N^{-1} \sum_i \dfrac{|e_i - \hat{e}_i|}{\min(e_i, \hat{e}_i)}$ |
| Prediction Performance | $pred_l = \dfrac{100}{N} \times \sum_{i=1}^{N} \begin{cases} 1 & if\ MRE_i \leq 0.25 \\ 0 & otherwise \end{cases}$ |

## 5. Results

This section presents performance figures of OABE against various adjustment techniques used in constructing ABE models. Since the selection of the best $k$ setting in OABE is dynamic, there was no need to pre-set the best $k$ value. In contrast, for other variants of adjustment techniques there was necessarily finding the best $k$ value that almost fits each model, therefore we applied different $k$ settings from 1 to 5 on each model as suggested by Li et al. [9] and the setting that produces best overall performance has been selected for comparison with other different models.

Table 4 *MMRE* and *Pred$_{25}$* Performance figures

| Dataset | MMRE | | | | | | Pred$_{25}$ | | | | | |
|---|---|---|---|---|---|---|---|---|---|---|---|---|
| | OABE | LSE | MLFE | RTM | GA | NN | OABE | LSE | MLFE | RTM | GA | NN |
| Albrecht | **40.2** | 62.9 | 65.2 | 61.2 | 45.4 | 51.2 | **44.6** | 37.5 | 37.5 | 33.3 | 33.3 | 29.2 |
| Kemerer | **39.6** | 41.4 | 64.5 | 44.6 | 60.4 | 166.0 | 53.3 | **60.0** | 26.7 | 33.3 | 33.3 | 13.3 |
| Desharnais | 34.5 | 37.2 | 45.6 | **33.4** | 49.4 | 78.4 | **48.2** | 42.9 | 37.7 | 41.6 | 37.7 | 31.2 |
| COCOMO | **50.1** | 65.8 | 148.2 | 54.0 | 159.5 | 203.6 | 20.2 | **31.7** | 14.3 | 25.4 | 14.3 | 6.3 |
| Maxwell | **41.7** | 71.2 | 71.2 | 46.4 | 117.2 | 199.9 | **34.4** | 27.4 | 27.4 | 32.3 | 17.7 | 3.2 |
| China | 24.7 | **20.9** | 32.8 | 36.5 | 46.5 | 68.6 | 80.7 | **82.4** | 25.9 | 45.9 | 43.9 | 46.1 |
| Telecom | **13.2** | 15.4 | 36.7 | 15.2 | 39.1 | 78.9 | **84.0** | 77.8 | 55.6 | 77.8 | 61.1 | 22.2 |
| Nasa | 61.2 | 58.3 | 55.7 | **54.9** | 58.6 | 99.2 | **50.0** | 33.3 | 33.3 | 33.3 | 38.9 | 11.1 |

Table 5 *MdMRE* Performance figures

| Dataset | MdMRE | | | | | |
|---|---|---|---|---|---|---|
| | OABE | LSE | MLFE | RTM | GA | NN |
| Albrecht | 37.2 | **29.7** | 30.3 | 40.5 | 38.5 | 43.1 |
| Kemerer | 23.3 | **21.3** | 39.6 | 46.1 | 41.4 | 128.5 |
| Desharnais | **26.3** | 28.9 | 31.0 | 30.9 | 35.9 | 51.9 |
| COCOMO | 47.7 | **38.0** | 71.6 | 46.9 | 81.1 | 99.5 |
| Maxwell | 44.2 | 48.1 | 48.1 | **41.0** | 60.2 | 160.0 |
| China | 24.6 | **22.6** | 84.4 | 28.4 | 29.2 | 29.2 |
| Telecom | **10.3** | 13.4 | 20.0 | 12.6 | 18.7 | 58.4 |
| Nasa | **25.8** | 39.4 | 44.1 | 36.6 | 31.5 | 81.3 |

Tables 4, 5 and 6 summarize the resulting performance figures for all investigated ABE models. The most successful method should have lower *MMRE*, *MdMRE*, *MMER*, *MBER* and higher *Pred$_{25}$*. The obtained results suggest that the OABE produced accurate predictions than other methods with quite good performance figures over most datasets.

Table 6 *MMER* and *MBRE* Performance figures

| Dataset | MMER | | | | | | MBRE | | | | | |
|---|---|---|---|---|---|---|---|---|---|---|---|---|
| | OABE | LSE | MLFE | RTM | GA | NN | OABE | LSE | MLFE | RTM | GA | NN |
| Albrecht | **38.6** | 57.2 | 50.0 | 86.1 | 53.1 | 154.4 | **61.2** | 87.7 | 82.7 | 107.5 | 65.8 | 166.0 |
| Kemerer | **51.3** | 59.7 | 55.5 | 53.8 | 56.8 | 73.3 | **57.5** | 71.4 | 83.9 | 64.8 | 81.1 | 124.3 |
| Desharnais | 37.2 | **35.2** | 38.0 | 40.7 | 47.4 | 95.1 | **40.4** | 45.6 | 54.1 | 46.8 | 65.5 | 81.4 |
| COCOMO | **58.0** | 62.9 | 226.6 | 117.8 | 285.2 | 111.9 | 97.3 | **92.9** | 319.4 | 129.0 | 383.3 | 239.4 |
| Maxwell | 54.7 | **48.3** | **48.3** | 63.1 | 108.2 | 117.4 | 84.2 | 81.9 | 81.9 | **74.3** | 175.9 | 199.8 |
| China | 16.2 | **14.8** | 47.1 | 55.2 | 44.8 | 64.4 | 23.3 | **23.0** | 32.1 | 62.1 | 62.3 | 90.1 |
| Telecom | **15.2** | 18.2 | 27.1 | 16.1 | 26.5 | 357.9 | **16.5** | 16.9 | 39.7 | 17.4 | 42.6 | 73.0 |
| Nasa | **44.4** | 49.3 | 53.0 | 80.5 | 46.6 | 279.4 | **71.1** | 75.6 | 73.7 | 98.0 | 74.1 | 99.6 |

However, these findings are indicative of the superiority of BA in optimizing $k$ analogies and adjusting the retrieved project efforts, and consequentially improve overall predictive performance of ABE. Also from the obtained results we can observe that there is evidence that using adjustment techniques can work better for datasets with discontinuities (e.g. Maxwell, Kemerer and COCOMO). Note that the result is exactly the "searching for the best $k$ setting" result as might be predicted by the researchers mentioned in the related work. We speculate that prior Software Engineering researchers who failed to find best $k$ setting, did not attempt to optimize this $k$ value with adjustment technique itself for each individual project before building the model.

Furthermore, two results worth some attention while we are carrying this experiment: Firstly, the general trend of predictive accuracy improvements across all error measures, overall datasets is not clear this certainly depends on the structure of the dataset. Secondly, there is no consistent results regarding the suitability of OABE for small datasets with categorical features (as in Maxwell and Kemerer datasets) but we can insist that OABE is still comparable to LSE in terms of *MMRE* and *Pred$_{25}$* and have potential to produce better estimates. In contrast, OABE showed better performance than LSE for the other two small datasets (NASA and Telecom) that do not have categorical features. To summarize the results we run the win-tie-loss algorithm to show the overall performance. Figure 3 shows the sum of win, tie and loss values for all models, over all datasets. Every model in Figure 2 is compared to other five models, over six error measures and eight datasets. Notice in Figure 2 that except the low performing model on, the tie values are in 49-136 band. Therefore, they would not be so informative as to differentiate the methods, so we consult win and loss statistics to tell us which model performs better over all datasets using different error measures. Apparently, there is significant difference between the best and worst models in terms of win and loss values (in the extreme case it is close to 119). The win-tie-loss results offer yet more evidence for the superiority of OABE over other adjustment techniques. Also the obtained win-tie-loss results confirmed that the predictions based on OABE model presented statistically significant but necessarily accurate estimations than other techniques. Two aspects of these results are worth commenting: 1) The NN was the big loser with bad performance for adjustment. 2) LSE technique performs better than MLFE which shows that using size measure only is more predictive than using all size related features.

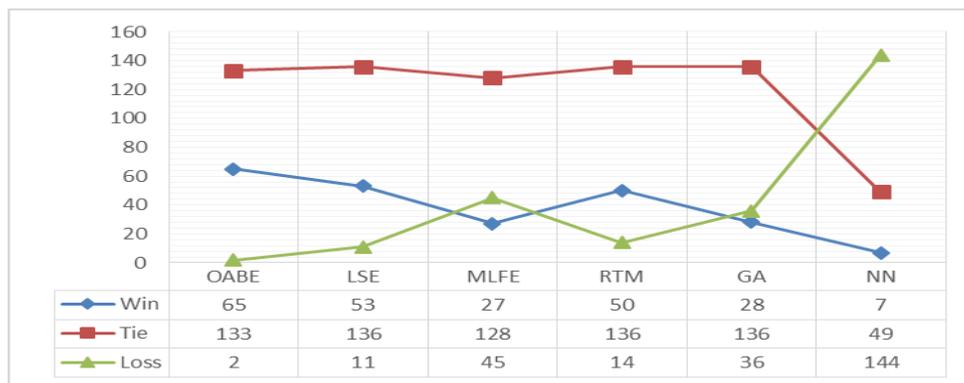

Figure 2. *win-tie-loss* results for all models.

We use the Bonferroni-Dunn test to compare the OABE method against other methods as shown in Figure 3. The plots have been obtained after applying ANOVA test followed by Bonferroni test. The ANOVA test results in *p*-value close to zero which implies that the difference between two methods are statistically significant based on AR measure. The horizontal axis in these figures corresponds to the average rank of methods based on AR. The dotted vertical lines in the figures indicate the critical difference at the 95% confidence level. Obviously, the OABE methods generated lower *AR* than other methods over most datasets except for small datasets. For such datasets, all models except NN generated relatively similar estimates but with preference to OABE that has smaller error. This indicates that OABE adjustment method is far less prone to incorrect estimates.

6. **Conclusions and Future Works**

This paper presents a new adjustment technique to tune ABE using Bees optimization algorithm. The BA was used to automatically find the appropriate *k* value and its feature weights in order to adjust the retrieved *k* closest analogies. The results obtained over 8 datasets showed significant improvements on prediction accuracy of ABE. We can notice that all models' ranking can change by some amount but OABE has relatively stable ranking according to all error measure as shown in Figure 2. Future work is planned to study the impact of using ensemble adjustment techniques.

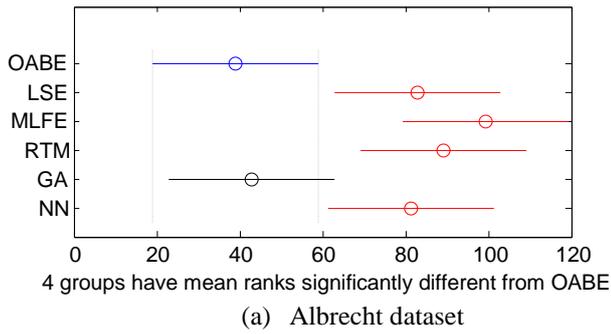
(a) Albrecht dataset

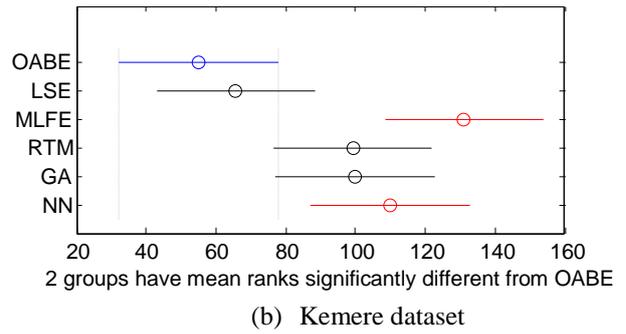
(b) Kemere dataset

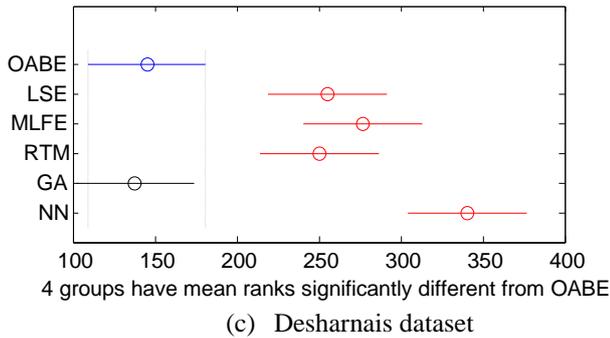
(c) Desharnais dataset

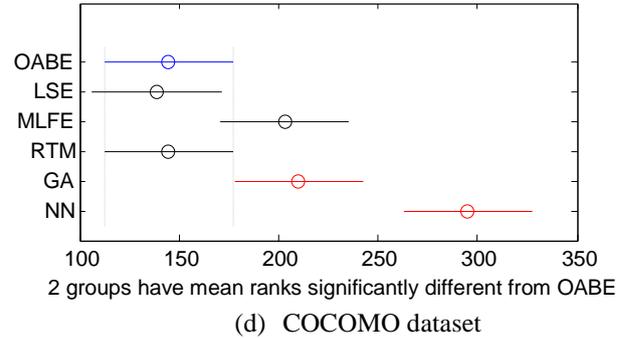
(d) COCOMO dataset

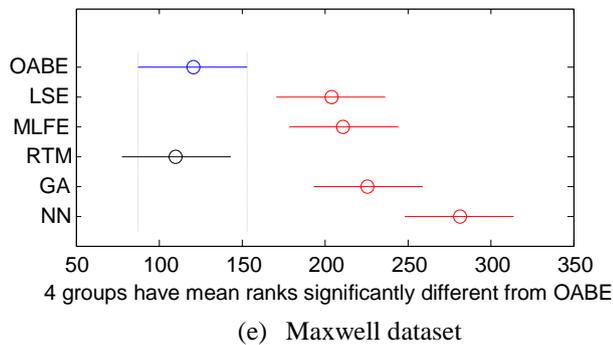
(e) Maxwell dataset

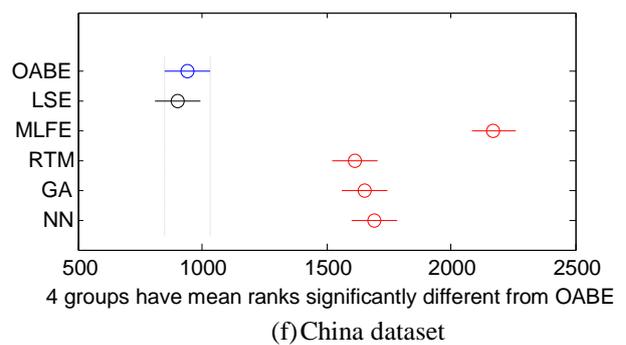
(f) China dataset

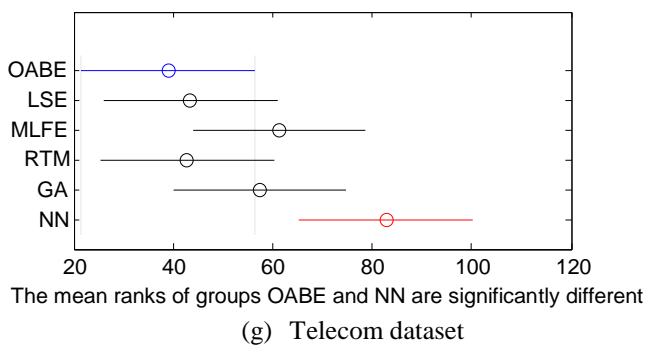
(g) Telecom dataset

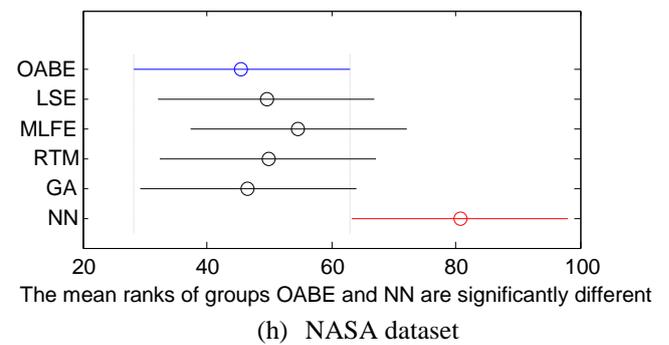
(h) NASA dataset

Figure 3. Bonferroni-Dunn test multiple comparison test over all datasets.

# 7. Acknowledgements

The authors are grateful to the Applied Science Private University, Amman, Jordan, for the financial support granted to this research project.